\documentclass{vgtc} 
\ifpdf
 \pdfoutput=1\relax          
 \usepackage{graphicx}        
 \DeclareGraphicsExtensions{.pdf,.png,.jpg,.jpeg} 
\else
 \usepackage{graphicx}        
 \DeclareGraphicsExtensions{.eps}   
\fi%

\graphicspath{{figures/}{pictures/}{images/}{./}} 

\usepackage{microtype}         
\PassOptionsToPackage{warn}{textcomp} 
\usepackage{textcomp}         
\usepackage{mathptmx}         
\usepackage{times}           
\usepackage{cite}           
\usepackage{tabu}           
\usepackage{booktabs}         
\usepackage[dvipsnames]{xcolor}
\usepackage{hyperref}

\onlineid{0}

\vgtccategory{SciVis}
\vgtcpapertype{application/design study}

\title{Explainable Spatial Clustering:\\Leveraging Spatial Data in Radiation Oncology}

\author{Andrew Wentzel\thanks{e-mail: awentzel@uic.edu}\\ %
    \scriptsize University of Illinois at Chicago\\ %
\and Guadalupe Canahuate\\ %
   \scriptsize University of Iowa
\and Lisanne V. van Dijk\\%
  \parbox{1.4in}{\scriptsize \centering University of Texas\\}
\and Abdallah S.R. Mohamed\\%
  \parbox{1.4in}{\scriptsize \centering  University of Texas\\}
\and C.David Fuller\\%
  \parbox{1.4in}{\scriptsize \centering  University of Texas}
\and G.Elisabeta Marai\\ %
   \scriptsize University of Illinois at Chicago} %

\shortauthortitle{Wentzel \MakeLowercase{\textit{et al.}}: Explainable Spatial Clustering}

\abstract{Advances in data collection in radiation therapy have led to an abundance of opportunities for applying data mining and machine learning techniques to promote new data-driven insights. In light of these advances, supporting collaboration between machine learning experts and clinicians is important for facilitating better development and adoption of these models. Although many medical use-cases rely on spatial data, where understanding and visualizing the underlying structure of the data is important, little is known about the interpretability of spatial clustering results by clinical audiences. In this work, we reflect on the design of visualizations for explaining novel approaches to clustering complex anatomical data from head and neck cancer patients. These visualizations were developed, through participatory design, for clinical audiences during a multi-year collaboration with radiation oncologists and statisticians. We distill this collaboration into a set of lessons learned for creating visual and explainable spatial clustering for clinical users.
} 

\keywords{Data Clustering and Aggregation, Life Sciences, Collaboration, Mixed Initiative Human-Machine Analysis, Guidelines}

\begin{document}

\firstsection{Introduction}

\maketitle
One of the most important applications of machine learning (ML) techniques to oncological healthcare is patient stratification. Stratification is the division of a patient population (group) into subgroups, or "strata". Each strata represents a particular section of that patient population. The strata are typically correlated with specific demographic or disease traits, and specific outcomes including survival or side effects in response to specific treatments. The nature of patient stratification makes it well suited for clustering---an unsupervised data mining technique that groups patients based on some measure of distance between them. When the distance measure and  clustering algorithm is well chosen, clustering can generate novel insights and help discover previously undiscovered structure in the data.

Oncological data is often tied to a patient's anatomy, which complicates the construction of a similarity measure between patients and the selection of a clustering algorithm. In cancer patients, the spatial information of the tumor and surrounding anatomy is vital in deciding optimal treatment and forecasting patient endpoints. Thus, understanding the underlying spatial structure of the data during the clustering process is important. Despite a widespread interest in sophisticated clustering techniques for patient stratification, the adoption of clustering in oncology is stifled by the difficulty in understanding the inner workings of spatially-informed clustering. 

In this work, we examine a participatory design of explanatory visual encodings born out of a long-term collaboration between oncology, data mining, and data visualization practitioners performing analysis on a cohort of head and neck cancer patients~\cite{marai2018precision,sheu2017conditional}. Specifically, this work looks at interpreting clusters of stratified head and neck cancer patients based on secondary disease spread to the lymph nodes, with the goal of helping clinical users understand the strata and use them to help predict the toxicity outcome of disease treatment. We reflect on the process of creating domain-specific visual encodings through participatory design to help "bridge the gap" between the data experts and healthcare experts~\cite{janicke2020participatory}. We further explore obstacles and successes when creating visual encodings for interpreting data mining techniques, and for communicating with oncology experts with limited background in both visualization and in artificial intelligence.

\section{Related Work}

{\bf Cluster Explainability }
Interpretation and visualization of clusters is a common analysis task tightly integrated with dimensionality reduction in general, but is less understood than traditional explainable AI (XAI) approaches, which are generally focused on supervised learning. A task analysis of 10 data analysts~\cite{brehmer2014visualizing} included 3 tasks related to clusters: verifying clusters, naming clusters, and matching clusters to existing classes. General methods of cluster visualizing have typically been linked to low-dimensionality embedding, where classes are shown plotted in a 2 or 3 dimensional space, and cluster-membership is shown on top of the data in the lower-dimension space~\cite{wenskovitch2017towards, alper2011design, efrat2014mapsets}. Hierarchical clustering methods, where clusters are iteratively created at different levels of granularity, have commonly been visualized as dendrograms. When dimensionality reduction isn't appropriate, general methods of multivariate data visualization are used, such as parallel coordinate plots~\cite{chou1999cluster} or specialized glyph encodings~\cite{cao2011dicon}. Other systems synthesize existing methods to support visual steering and clustering for scientists~\cite{cavallo2018clustrophile, metsalu2015clustvis, cava2017clustervis}. While some recent work has dealt with clustering ensemble geospatial data~\cite{ma2018interactive}, we are not aware of any methods that deal explicitly with clustering anatomical or 3-d data as in this work.

	
\noindent{\bf Vis in Healthcare}
Visualization approaches to healthcare problems often focus on supporting data exploration, rather than understanding predictive models~\cite{bernard2018using, bui2007timeline, loorak2015timespan}. Certain systems for model exploration have been developed to aid in the development of regression models based on the workflows of biostatisticians~\cite{dingen2018regressionexplorer, rojo2020gacovi}. Other systems have applied visualization for clustering cancer data~\cite{metsalu2015clustvis}, and predicting infection spread in hospital wards~\cite{muller2020visual}. For spatial data, Grossmann et al.~\cite{grossmann2019pelvis} incorporated methods for visualizing clusters based on bladder shape to support a retrospective study on prostate cancer patients. Some works have attempted to identify design considerations when working with domain experts in healthcare~\cite{raidou2020lessons, maack2020towards}. However, with the exception of Raidou et al. ~\cite{raidou2020lessons}, most of these considerations do not apply to clustering or spatial data, and are largely focused on analytics and electronic health record data. As a result, there is a dearth of papers discussing how to approach unsupervised XAI models to reach clinical audiences.

\section{Background}
In many cancer patients, tumors metastasize into the lymphatic system, causing lymph nodes to become "involved"---affected by secondary nodal tumors. The lymphatic system forms a complex chain of lymph nodes, and these secondary tumors spread along these chains to adjacent regions stochastically. Affected lymph nodes are a long-established factor in determining patient outcomes in head and neck cancer~\cite{leemans1994recurrence}. Current predictive systems use a staging system based on the size and number of nodal tumors, but miss more nuanced predictions about how the different patterns of nodal spread may affect toxicity outcomes~\cite{wu2019integrating, huang2017overview}. No prior machine learning methods correctly handle this type of spatial data, due to a lack of spatial similarity measures~\cite{luciani2020spatial, elhalawani2018machine}.

Our data comes from a cohort of 582 head and neck cancer patients collected retrospectively from the MD Anderson Cancer Center. All patients survived for at least 6 months after treatment. Data was collected on the presence of 2 severe side effects: feeding tube dependency, and aspiration - fluid in the lungs that requires removal. We mainly consider the presence of either of these side effects, which we define as as radiation-associated dysphagia (RAD)~\cite{christopherson2019chronic}. The data also encodes the disease spread to 9 connected regions (denoted as levels 1A-6) on each side of the head, along with the disconnected retropharyngeal lymph node (RP or RPN). Many patients in this cohort had unique patterns of disease spread to the lymph nodes. 
	
The project consisted of 2 phases with distinct design requirements. In phase 1 (model development), we worked alongside six domain experts in radiation oncology, and two data analysts with data mining and biostatistics backgrounds, over four years. During this time we developed, validated, and deployed an anatomically-informed patient stratification method based on each patient's patterns of diseased lymph nodes~\cite{luciani2020spatial}. To demonstrate the important role of spatiality, the stratification used only anatomical features. We met with representatives from this group up to three times per week via teleconferencing, as well as in quarterly face to face meetings. In phase 2 (model dissemination), our results needed to be analyzed and delivered to the larger radiation oncology community. In this stage, we received feedback from three additional radiation oncologists and two bioinformaticians with expertise in head and neck cancer. The final stratification approach is available to clinicians through an open-source interface~\cite{marai2018precision}. Below, we reflect on the design process, which focused on an activity-centered design paradigm~\cite{marai2017activity}, along with feedback from the domain experts.

\section{Model Development Phase}
\begin{figure}[ht]
  \includegraphics[width=\columnwidth]{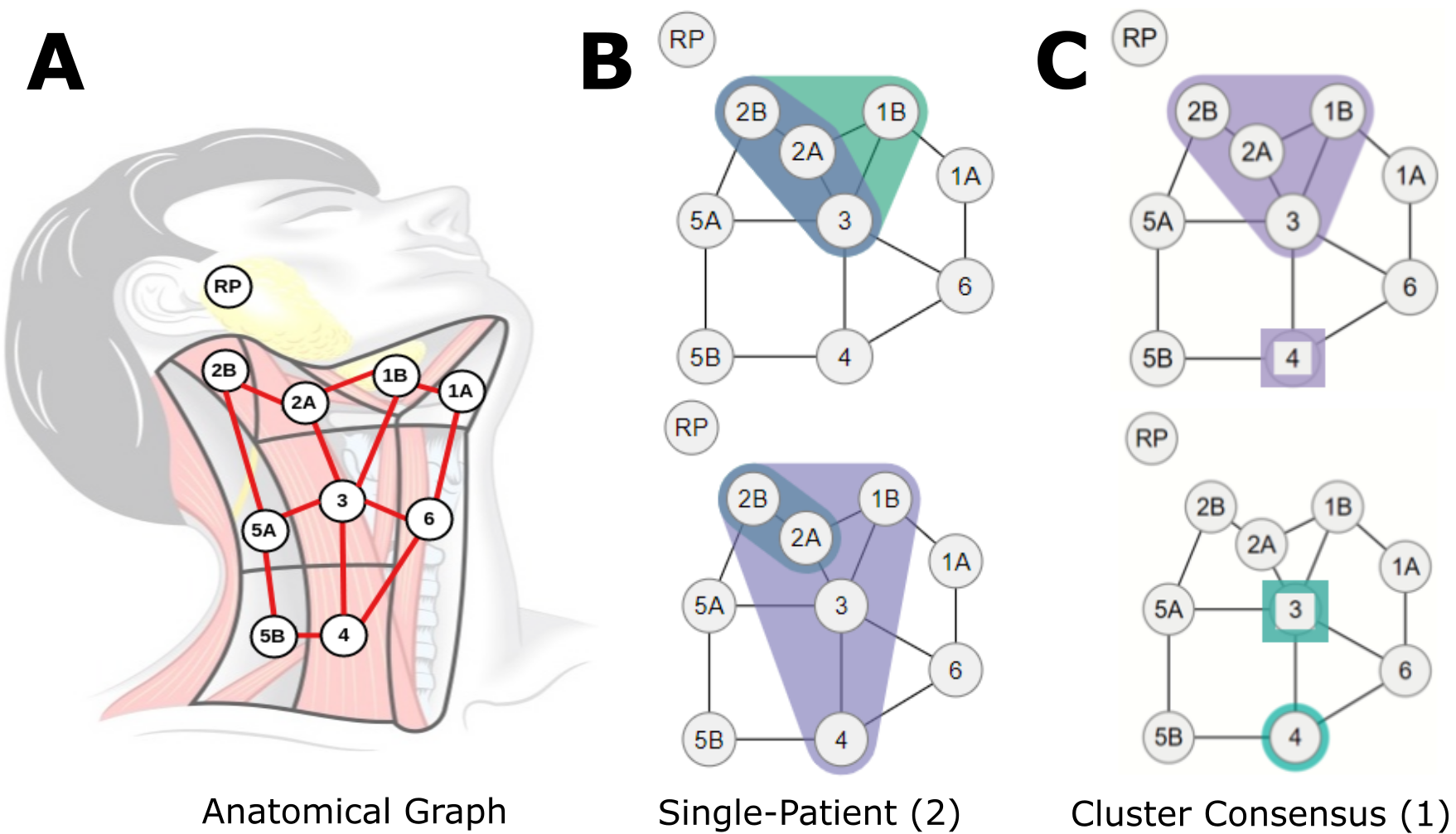}
  \caption{(A) Lymph nodes overlaid over a diagram of the neck. (B) Example graphs of diseased nodes for 2 individual patients (datapoint representation). (C) Example consensus graph for 1 cluster (cluster representation). The top-right graph shows disease spread with 66+\% of patients on the right nodes in 1B, 2A, 2B, and 3, and disease in 1-33\% of patients in right node 4. The bottom-right graph similarly indicates involvement of $>$66\% and $<$33\% of patients in left nodes 4 and 3, respectively.}
  \label{fig:graph}
\end{figure}

\begin{figure}[ht]
  \includegraphics[width=\columnwidth]{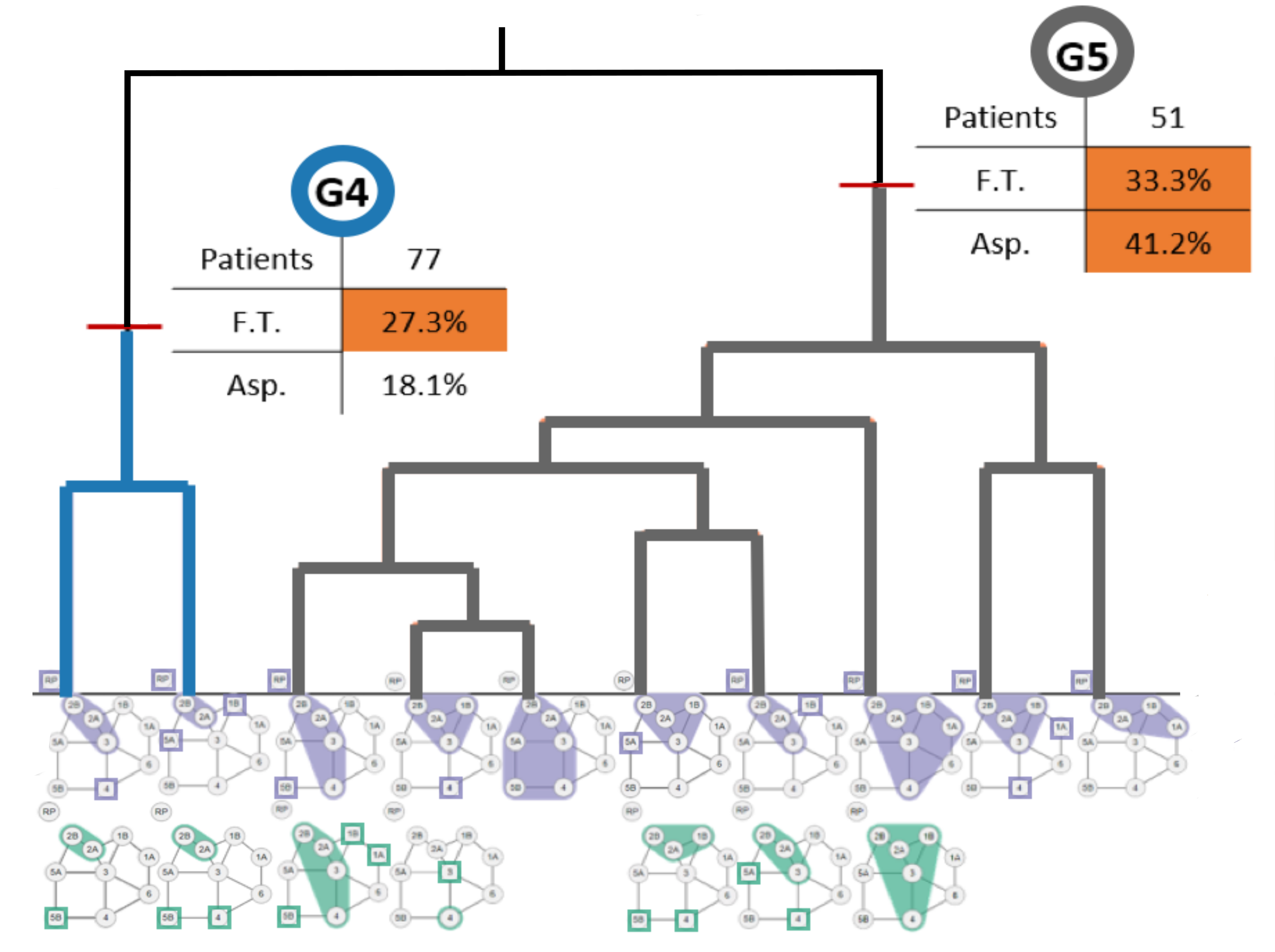}
  \caption{Part of an augmented dendrogram of lymph node clusters (clusters 1-3 not shown; the full dendrogram is available in \href{http://www.sciencedirect.com/science/article/pii/S2590177X20300019\#f0050}{Luciani et al.} Leaves of the tree are smaller clusters that merge at higher levels according to the agglomerative clustering algorithm. Clusters are id-ed by colors in the graph. Clusters are further augmented with breakdowns of relevant clinical covariates of interest (F.T.: Feeding Tube; Asp.: Aspiration).}
  \label{fig:dendrogram}
\end{figure}

\begin{figure}[ht]
  \includegraphics[width=\columnwidth]{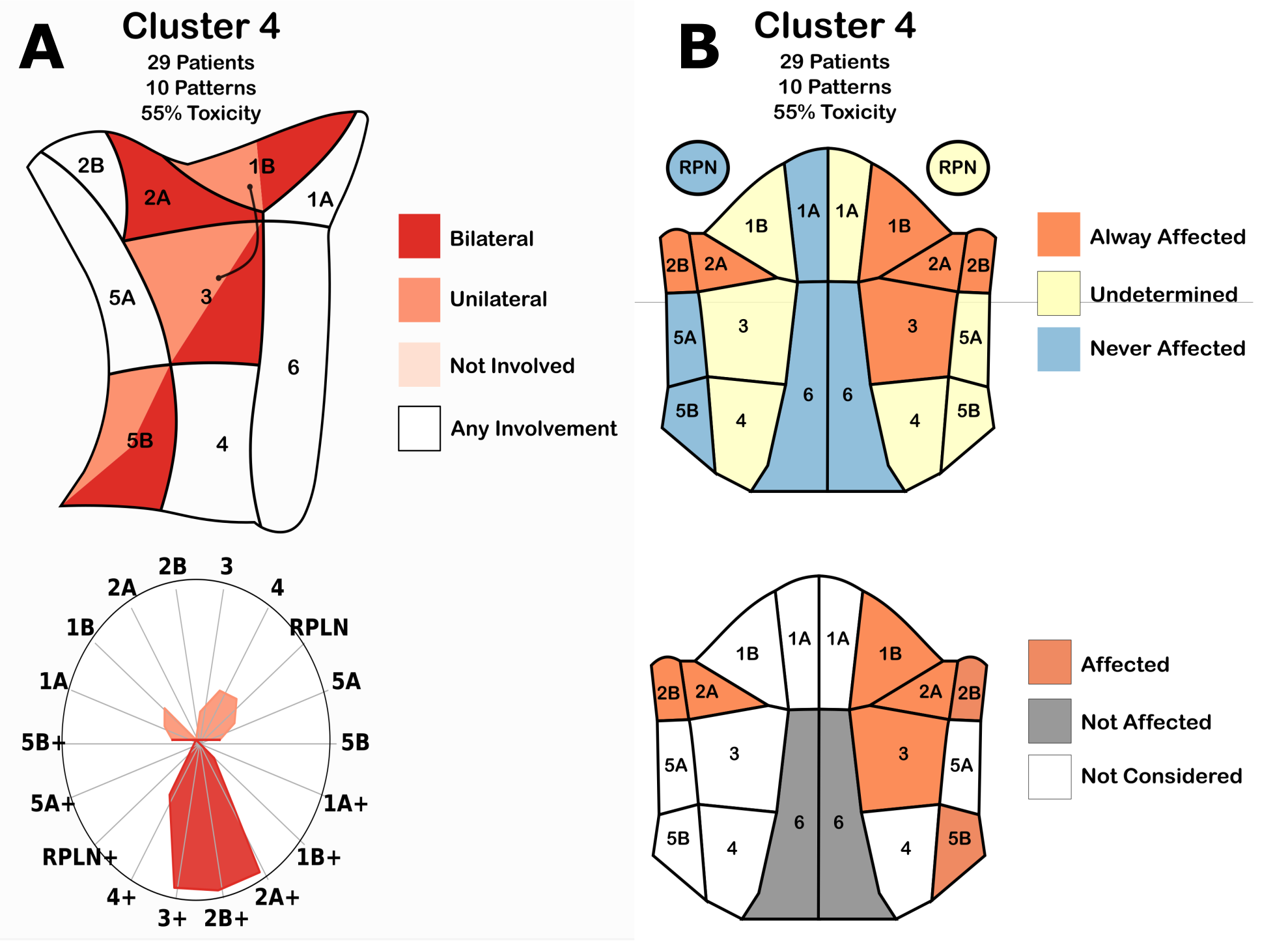}
  \caption{(A) Cluster conditionals. (Top-left) Map of the regions in the neck. Color indicates when the decision tree classified a patient into the cluster based on if the region had no disease (pale red), tumors in one side of the head (red), both sides of the head (dark red), or a combination of two options. (Bottom-left) Radar chart showing the percentage of patients in the cluster with nodal tumors in a given region. Color indicates the presence of tumors in exactly one (pale red) or two (dark red) sides of the head. (B) Second iteration of cluster conditionals. (Top-right) Membership diagram showing the regions in the head. Color indicates when all (red), a subset of (yellow), or none of (blue) the patients in a cluster had nodal tumors in a region. (Bottom-right) Decision-tree based diagram. Colors indicate when a decision tree classified a patient into that cluster. }
  \label{fig:experimental}
\end{figure}

\begin{figure}[ht]
  \includegraphics[width=\columnwidth]{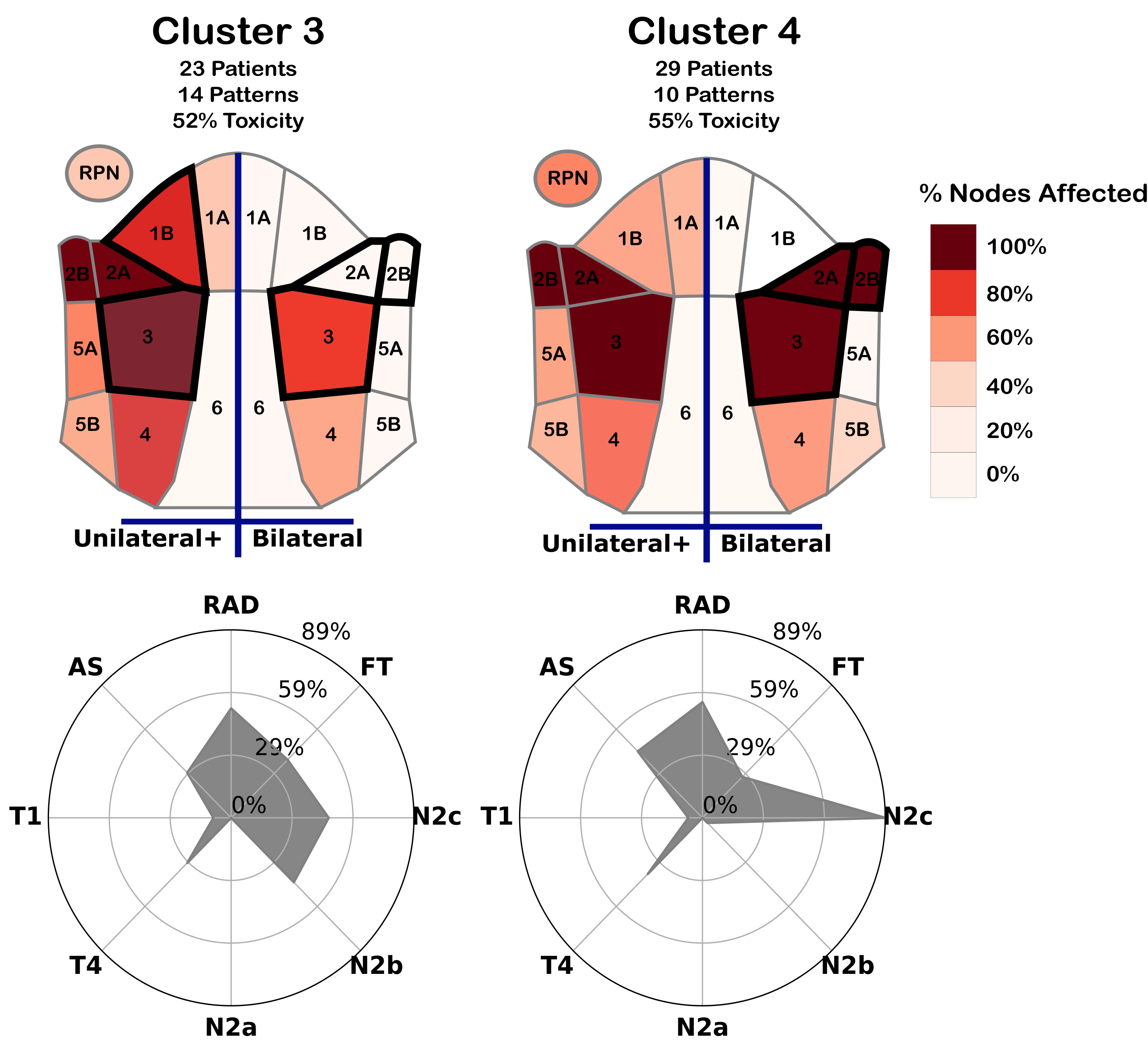}
  \caption{Designs for two high-risk cluster conditionals. (Top) Spatial heatmaps showing the portion of patients with nodal tumors in each region for at least one (left) or both (right) sides of the head. Regions most informative in determining cluster membership are outlined in a thick dark border. (Bottom) Radar charts showing the percentage of patients within the cluster with a given toxicity outcome (FT/RAD/AS), and those within an existing risk-staging group (T1/T4/N2a/N2b/N2c).}
  \label{fig:heatmap}
\end{figure}

In phase 1, we worked to identify a meaningful, anatomically-informed distance measure between patients, as well as an appropriate method of clustering the patients. We developed an approach in which each side of the head was treated as a graph. Nodes in this graph corresponded with regions in the head that aligned with those used in existing oncology literature, and regions that were anatomically adjacent in the head were connected in the graph as an edge. Each patient was treated as two sub-graphs, one for each side of the head, containing only the nodes with nodal tumors. A distance measure based on these graphs then needed to be identified, alongside a clustering technique that led to meaningful clusters (activity 2). Clustering was performed using only the spatial disease spread captured by the graph model. Because identifying relevant structures in oncological data is nontrivial, defining this methodology required iterative experimentation with different features, clustering techniques, numbers of clusters, and other parameters~\cite{tosado2020clustering}. We identified the following activities that required visual support:

\begin{enumerate}
  \item Identify and analyze the relevant spatial data features underlying one datapoint (i.e. patient).
  \item Analyze the effects of different spatial similarity measures on clustering (i.e. why two patients are considered to be similar under a specific measure).
  \item Analyze the representative patterns and pattern variation within each cluster.
\end{enumerate}


\paragraph{Datapoint Representation}
The first design followed a graph metaphor to encode the diseased regions for each patient (activity 1). A compact graph that followed an anatomical map of lymph node chains for half the head (because the problem is symmetric) was used as a template for each patient (Figure~\ref{fig:graph}-A), based on ideas from biological network visualization~\cite{marai2019ten, wenskovitch2014mosbie}. For each patient, two envelopes were drawn over their diseased nodes. Green and purple envelopes were used for the left and right side of the head, respectively. Areas where envelopes overlap are shown in blue and denote regions where tumors occur on both sides of the head, which are of particular interest to oncologists (Figure~\ref{fig:graph}-B).
 
This design allowed for a compact representation of a complex spatial feature space, while following the mathematical intuition behind different distance measures. These graphs were incorporated into an interface that shows patients and compares them to their most similar matches. The compact representation was useful in identifying the spatial features of each datapoint, as well as interpreting distance between patients.

\paragraph{Cluster Representation}
In a first attempt to characterize each cluster, we selected a representative patient for each cluster: i.e., the patient closest to the cluster centroid (activity 3). The representative patient, however, did not capture any intra-cluster variability. Subsequently, we created a new representative encoding by placing the most commonly affected nodes for a cluster in a "consensus" graph. Nodes where $\frac{2}{3}$ of the patients in that cluster had nodal tumors were outlined in envelopes. However, in this new representation it was unclear why certain clusters were not merged. In a third iteration, we added a different marker (squares) for nodes where less than $\frac{2}{3}$ of the patients in that cluster, but at least one patient had nodal tumors (Figure~\ref{fig:graph}-C). We used shape, rather than color, because hue already encoded disease laterality, and further intensity variation was not legible given the small scale.

However, at small scale, the markers and colors for multiple clusters became hard to distinguish. Additionally, outside clinicians and bioinformaticians mis-interpreted the third encoding as representing only one patient in that cluster, and in one case, as clusters containing identical patients. In the fourth design, two stacked graphs were used for each side of the head for each cluster, and visual scaffolding~\cite{marai2015visual} was used to explain the progression from a single datapoint representation to the consensus graph. The consensus graphs were placed within dendrograms, which showed the consensus graphs of smaller component clusters within each larger cluster of interest (Figure~\ref{fig:dendrogram}). To further clarify the hierarchical clustering process, we added explicit color-coding of the dendrograms, with labels and colors showing the cluster names and tracing the merging process, as well as small statistics tables showing the patient toxicity outcomes within each larger cluster.

\section{Clinical Model Dissemination Phase}
In the second phase, our results needed to be able to reach their intended audience: clinical radiation oncologists. While the methodological development was concerned with the clinical validity of the analysis, clinical readers are more concerned with significance of the results, and place more importance on feasibility, trust in the underlying covariates, and the implications of the results~\cite{wentzel2019cohort, wentzel2020precision}, rather than the methodology used, which had already been peer-reviewed~\cite{luciani2020spatial}. In this phase, we used four clusters to align with existing staging systems, and the clustering still only considered spatial disease spread. In order to effectively communicate results, we identified the following activities to support:
\begin{enumerate}
  \item Describe patient clusters from an anatomical perspective.
  \item Identify each cluster's underlying structure.
  \item Connect structural cluster differences to clinical covariates.
  \item Explain plausible causal relationships between the clusters and correlated patient outcomes.
\end{enumerate}

\paragraph{Cluster Conditionals} 
The first design relied on two synergistic encodings for each cluster. The first encoding expanded on the original anatomical diagram to show the most discriminative features in each cluster (conditionals). To do this, a decision tree was trained on the cohort to predict cluster membership with 100\% accuracy using the number of sides of the head with a nodal tumor in each region of the head and neck, which could be 0 (no disease), 1 (unilateral disease), or 2 (bilateral disease). Because experts who had not participated in the methodology design process had trouble understanding the graph-based encoding, the set of variables considered sufficient to any patient in the training data into a given cluster was then encoded into an anatomical region diagram of one side of the neck (Figure~\ref{fig:experimental}-A). By focusing on the regions that the decision tree considered, the diagram highlighted the regions that best identified the key differences between clusters, while omitting regions with commonalities between then, in order to support activities 2 and 3. The second encoding was a radar plot of the percentage of people in a cluster with either unilateral or bilateral disease spread in a given region of the neck. This representation allowed for a more detailed view of the overall distribution of tumors in each cluster (activity 1).

The initial cluster visualization design using trees was found to intuitively make sense to clinical collaborators. However, they had difficulty understanding the underlying explanation of the diagrams and how they were generated within the space of a figure caption, as they had limited experience with decision trees. Collaborators incorrectly assumed that all combinations of nodal disease in the diagrams were shared between all patients in a given cluster. Additionally, our collaborators pointed out that while the one-sided diagram of the neck was common for surgical applications, radiation oncologists often visualized the neck in terms of a front view that included both sides of the head simultaneously. 

In the second design (Figure~\ref{fig:experimental}-B), each cluster is encoded using a frontal view anatomical diagram. A red-yellow-blue categorical color scheme was used to mark which regions were diseased in all patients, some patients, or no patients within the cluster, respectively, following the original intuition of our collaborators. An additional anatomical diagram based on the decision tree was included for each cluster below the membership diagrams. Since the new diagram included both sides of the head, color was used to show when the decision tree split the cluster based on the presence of disease (red), or absence of disease (gray) in a given region, while white regions were not considered in the model.

\paragraph{Cluster Membership}
The conditional designs were better-received by the clinicians, but difficulties in understanding the colormap and the lack of detail in the cluster membership made it challenging to correctly draw insights. To address these concerns, we designed a new heatmap diagram of the neck (Figure ~\ref{fig:heatmap}), which used a sequential white-red color scheme to encode the number of patients in a cluster with disease in a given region (activity 1). We note that head and neck oncologists account for symmetry when discussing similar patients, and thus a symmetric encoding was a desired feature. A simplified decision tree was trained to identify the regions that contained the most information about cluster membership, which were outlined with a dark border in the heatmaps (activity 2). Additional labels were included, to indicate the left/right sides of the diagram show unilateral vs. bilateral involvement, rather than the literal left/right sides of the head.

To help indicate the relationship between the clusters and other clinical data, covariates and outcomes that were the most interesting to clinicians were included in a radar chart alongside the heatmaps for each cluster. The inclusion of these data helped with the collaborators' ability to discuss potential relationships between the structure of the clusters and correlated outcomes (activities 3 and 4). 

\section{Design Lessons}
Through the course of these iterations, we have distilled design lessons for interpretable clustering with spatial data. 

\noindent{\bf L1. }\textit{Use visual scaffolding based on users' spatial background.} Spatial representations were, as expected, essential to understanding the clustering. Furthermore, encodings were better received when they mapped directly to the users' model of the problem, particularly when the users did not participate in the design. Using a graph-based encoding for the patient lymph node chains allowed us to draw parallels to graph theory, which was useful when testing similarity measures that were based on graph matching methods. In contrast, when designing for the wider oncology community, the encoding best received was created by visually scaffolding the graph directly onto an anatomical diagram of the neck from clinical literature. 

\noindent{\bf L2. }\textit{Incorporate visual details specific to the user's activities.} When designing for the methodology development, we focused on developing the clustering algorithm and ensuring that the results were more meaningful than existing methods. Placing the cluster visualizations within a dendrogram allowed the users to scrutinize the inner workings of the clusters at different scales. In contrast, clinicians were more results-focused. Namely, their key interests focused on the spatial structure underlying each cluster, how the clusters related to outcomes and existing clinical categories, and if these correlations could be explained in a way that was supported by clinical intuition. Thus, the design benefited from incorporating anatomical details and additional clinical covariates that were not considered when designing the model.

\noindent{\bf L3. }\textit{Show secondary variables and outcomes.} Design iterations that failed to include explicit labeling of results directly into the figure led to confusion. In the initial dendrograms, viewers had trouble connecting the clusters directly to other statistical analysis. For the clinical figures, collaborators often assumed that there were direct causal relationships between variables shown in the figure. In this case, it was useful to include potential confounding variables, to allow the readers to come up with alternative hypotheses.

\noindent{\bf L4. }\textit{Design for both interactive and static visualization.} In our experience, we started out with interactive designs aiming to assist a relatively small group of domain experts, who participated in the design process. Relatively quickly, it became obvious that the spatial clustering had to be explained to a broader audience that expected static visualizations, in the style of biomedical illustrations. Future works will stay closer to the illustrative style during the interactive model development phase, to reduce the cost of later redesign.

\noindent{\bf L5. }\textit{Build decision trees and conditionals to help explain spatial cluster differences.} When working with the broader audience, we found that the easiest way to explain cluster differences required explicit construction of decision trees, and "conditionals" based on the structure of the data---attempting to directly encode the differences was infeasible.

\section{Conclusion}
This work reflects on the process of designing visualizations for clustering with anatomical spatial data. These designs were developed in two phases over several years, using participatory design alongside collaborators with background in bioinformatics and radiation oncology. Through these designs iterations, we distill a set of lessons learned. While we focus on a particular problem, our design approach can be generalized to other type of cancer with spatially dependent data. These designs are part of a larger body of work borne out of a multi-year collaboration with domain experts with anatomical cancer data. By incorporating additional insights from sibling projects, we aim to develop a comprehensive set of design guidelines for visualizing clusters of spatial data and effectively disseminating these results to domain expert audiences outside of the visualization community.

\acknowledgments{This work is supported by the US National Institutes of Health, through awards NIH NCI-R01CA214825 and NIH NCI-R01CA2251. We thank all members of the Electronic Visualization Laboratory, members of the MD Anderson Head and Neck Cancer Quantitative Imaging Collaborative Group, and our collaborators at the University of Iowa and University of Minnesota.}

\bibliographystyle{abbrv-doi}

\bibliography{template}
\end{document}